# FILTRATION OF MICROPOLAR LIQUID THROUGH A MEMBRANE COMPOSED OF SPHERICAL CELLS WITH POROUS LAYER

D. Yu. Khanukaeva

*Gubkin Russian State University of Oil and Gas (National Research University)*
*Leninsky prospect, 65-1, Moscow, 119991, Russia*
*e-mail: khanuk@yandex.ru*

**Abstract:** This paper considers membranes of globular structure in the framework of the cell model technique. Coupled micropolar and Brinkman-type equations are used to model the flow of micropolar fluid through a spherical cell, consisting of solid core, porous layer and liquid envelope. The solution is obtained in analytical form. Boundary value problems with different conditions on hypothetical cell surface are considered and compared. The hydrodynamic permeability of a membrane is investigated as a function of micropolar and porous medium characteristics.

**Keywords:** micropolar fluid flow; cell model; porous medium; hydrodynamic permeability

**Acknowledgement:** The present work is supported by Russian Foundation for Basic Research (19-08-00058).

## Introduction

The majority of porous media, especially some types of membranes, can be represented as a chaotic assemblage of particles of various shapes and sizes [1,2]. Fibrous membranes are usually modeled as a package of cylindrical fibers or cylindrical cells. Globular structures can be described by a swarm of spherical or spheroidal particles of some average size, defined by the data on pore size distribution.

The Happel-Brenner cell model [3] is widely used for the modeling of filtration flows in porous structures. For flows of Newtonian liquids, this method is well developed for all types of geometrically symmetric cells [4-14]. The idea is to consider a single particle encapsulated in a hypothetical cell; the effect of neighboring particles is taken into account via boundary conditions on the cell surface. The particle can be solid, porous, or have a solid core covered by a porous layer. Alternatively two immiscible liquids can taken to be the core and the envelope respectively. A cell consisting of a solid core and porous non-deformable hydrodynamically uniform layer, that simulates a partially degraded membrane, was originally developed in [5]. The Stokes-Brinkman system of governing equations was solved with various types of boundary conditions on solid surfaces and outer hypothetical cell surface, namely, Happel [15, 16], Kuwabara [17], Mehta-Morse [18] or Cunningham [19], Kvashnin [20] conditions.

All of the papers mentioned above have considered only Newtonian fluids, while a great amount of liquids exhibit rheology and properties (especially, on micro scales and in the vicinity of boundaries), which cannot be adequately described by classical models. The existence of a wide variety of non-Newtonian models confirms the fact that all of them are far from being universal, though in some particular cases very good agreement with observations has been achieved. The situation is even worse with flows in porous media, as mentioned in review [21].

Filtration of liquids with microstructure has remained almost unstudied, while the theory and applications of free micropolar flows have been well developed. Media with microstructure introduced by the Cosserat brothers [22] consist of elements, which can rotate independently of their translational motion. They experience stresses and couple stresses described by the non symmetric tensors. Therefore such liquids are called polar or micropolar to distinguish them from Newtonian liquids, which are non-polar media. For simple micro fluids the dependency between the deformation rate tensor and the stress tensor remains linear in contrast to non-Newtonian models. Also, the curvature-twist rate tensor is linearly related to the couple stress tensor for simple micro fluids. The mathematical theory of micropolar flows was offered by Eringen [23, 24] and has been actively developed in the last decades. The application of the micropolar fluid theory includes but is not restricted to the flows of suspensions, lubricants, physiological liquids such as blood and synovial liquid. Besides, many problem statements in the framework of simple microfluids allow analytical solutions both for free and for filtration flows. A review of the existing analytical solutions and basic applications of simple microfluids was given by Khanukaeva and Filippov [25]. Also, the formulation of boundary value problems for composite cells with a solid core, porous layer and micropolar liquid layer was presented in the aforementioned review.

So, the micropolar model seems to be very efficient for simulating dispersed media flows in porous regions.

Only few works dealt with micropolar flows in cell models [26-29]. The problem of flow along the axis of composite solid-porous cylindrical cell was solved and analyzed in [30]. Perpendicular flow in a cylindrical cell of the same structure was considered by the same authors in [31]. To the best of our knowledge, a combined solid-porous particle in a spherical cell has not been studied before. The present paper is devoted to the modeling of a micropolar liquid flow through a membrane, which is represented by a set of spherical cells. All known types of boundary conditions at the outer surface of the cell are considered and compared, as thus far none of them turned out to be more preferable than the others. The appropriate generalization of the boundary value problem for micropolar liquids is also made and discussed.

The obtained analytical solution of the problem is used for the calculation of the hydrodynamic permeability of the membrane as a whole. Owing to the explicit form of the obtained expression for the hydrodynamic permeability its dependencies on the boundary conditions, liquid and porous layer characteristics can be investigated in their whole ranges. However, cumbersome representation of the expression does not allow its analytical study. Therefore a parametric investigation of the hydrodynamic permeability is fulfilled using machine computational capacities. A comparison with the experimental data is also presented.

**1. Statement of the problem**

The cell consists of three concentric layers as it is shown in Fig.1. The solid core has radius $a$, it is covered with a uniform porous layer $a < r < b$ (Region 1), which in turn is surrounded by a free micropolar liquid layer $b < r < c$ (Region 2). The spherical coordinate system $(r, \theta, \varphi)$ is introduced so that the direction of uniform flow velocity vector **U** corresponds to $\theta = 0$, $(0 \leq \theta \leq \pi, \ 0 \leq \varphi < 2\pi)$. The velocity magnitude is small enough for the Stokes approach to be valid.

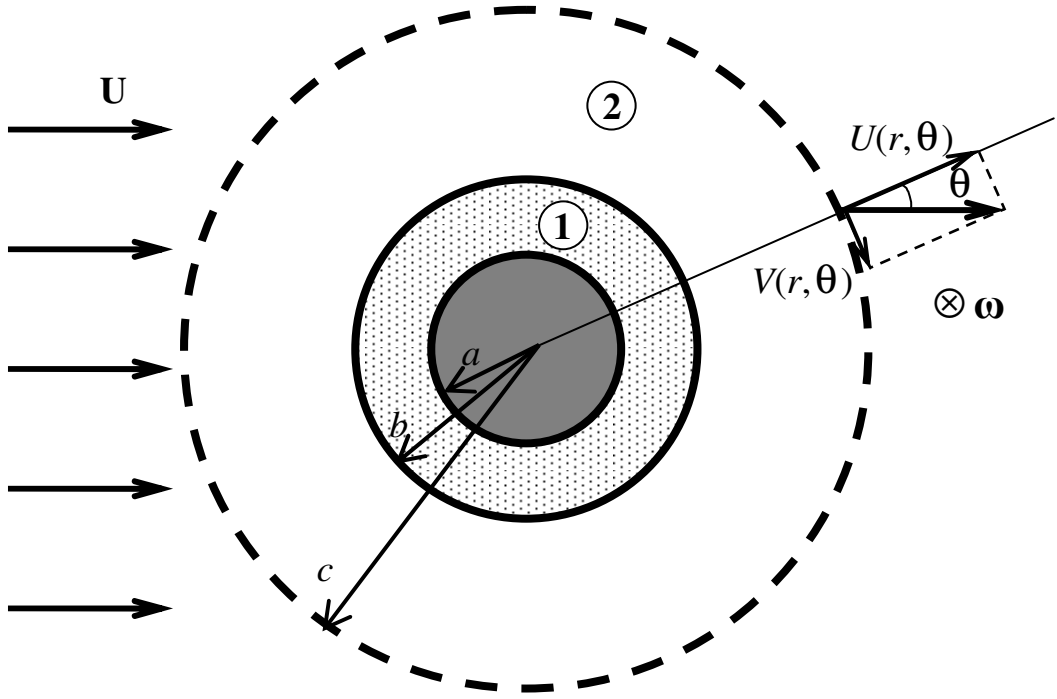

**Fig. 1** The scheme of the flow

According to the theory of micropolar fluids, developed by Eringen [24, 32], the steady creeping flow in Region 2 is governed by the following equations: the continuity equation, the momentum equation and the moment of momentum equation

$$\nabla \cdot \mathbf{v} = 0,$$
$$\mathbf{0} = \rho \mathbf{F} - \nabla P + (\mu + \kappa)\Delta \mathbf{v} + 2\kappa \nabla \times \boldsymbol{\omega},$$
$$\mathbf{0} = \rho \mathbf{L} + (\alpha + \delta - \varsigma)\nabla\nabla \cdot \boldsymbol{\omega} + (\delta + \varsigma)\Delta\boldsymbol{\omega} + 2\kappa \nabla \times \mathbf{v} - 4\kappa\boldsymbol{\omega},$$

where $\mathbf{v}, \boldsymbol{\omega}$ are the linear velocity and angular velocity (or spin) vectors correspondingly, $P$ is the pressure, $\rho$ is the liquid density, $\mathbf{F}, \mathbf{L}$ are the body force and body moment per unit mass respectively, $\mu, \kappa, \alpha, \delta, \varsigma$ are the viscosity coefficients of the micropolar medium.

Coefficient $\mu$ is an ordinary coefficient of dynamic viscosity for the Newtonian liquid, obtained from the considered micropolar liquid in the limiting case of zero rotational viscosity $\kappa$. In the theory of simple micro fluids these coefficients linearly relate the stress tensor $\hat{t}$ to the symmetric $\hat{\gamma}^{(S)}$ and skew symmetric $\hat{\gamma}^{(A)}$ parts of the deformation rate tensor $\hat{\gamma}$, defined as $\hat{\gamma} = (\nabla \mathbf{v})^T - \hat{\varepsilon} \cdot \boldsymbol{\omega}$, where $\hat{\varepsilon}$ is the Levi-Civita tensor. The stress tensor is presented as a sum of the symmetric and skew symmetric parts, the spherical part being written separately, as it is usually done in classical hydrodynamics:

$$\hat{t} = (-P + \lambda \operatorname{tr}\hat{\gamma})\hat{G} + 2\mu\hat{\gamma}^{(S)} + 2\kappa\hat{\gamma}^{(A)},$$

where $\hat{G}$ is the metric tensor. One can note that $\operatorname{tr}\hat{\gamma} = 0$ for incompressible fluids, so consideration of the coefficient $\lambda$ may be omitted.

Angular viscosities $\alpha, \delta, \varsigma$ are the coefficients in the constitutive equation respectively relating the spherical, symmetric and skew symmetric parts of the curvature-twist rate tensor $\hat{\chi} = (\nabla\boldsymbol{\omega})^T$ with the couple stress tensor $\hat{m}$:

$$\hat{m} = \alpha(\operatorname{tr}\hat{\chi})\hat{G} + 2\delta\hat{\chi}^{(S)} + 2\varsigma\hat{\chi}^{(A)}.$$

The chosen notation of the viscosity coefficients follows the form accepted in the micropolar theory of elasticity [33] and slightly differs from the original notation of Eringen. The only reason for this choice is the convenience in comparison with the non-polar limiting case. The correspondence with the original notation of [24, 32] can be achieved by formal re-notation.

Using the equality $\nabla \times \nabla \times \mathbf{a} = \nabla\nabla \cdot \mathbf{a} - \Delta\mathbf{a}$, the continuity equation and $\nabla \cdot \boldsymbol{\omega} = 0$ arising from the symmetry of the cell, the governing equations for the free micropolar fluid in the absence of external forces and couples can be written as ($b < r < c$)

$$\begin{aligned} &\nabla \cdot \mathbf{v}_2 = 0, \\ &-(\mu + \kappa)\nabla \times \nabla \times \mathbf{v}_2 + 2\kappa\nabla \times \boldsymbol{\omega}_2 = \nabla P_2, \\ &-(\delta + \varsigma)\nabla \times \nabla \times \boldsymbol{\omega}_2 + 2\kappa\nabla \times \mathbf{v}_2 - 4\kappa\boldsymbol{\omega}_2 = 0, \end{aligned} \quad (1)$$

where subscript 2 corresponds to the free stream layer 2 in Fig.1. Subscript 1 will be used for the porous Region 1, where the filtration flow takes place.

The filtration flow of micropolar liquid is governed by the Brinkman-type equations, derived by Kamel et al. [34] using the intrinsic volume averaging technique:

$$\begin{aligned} &\nabla \cdot \mathbf{v} = 0, \\ &\nabla P = \left(\frac{\mu}{\varepsilon} + \frac{\kappa}{\varepsilon}\right)\Delta\mathbf{v} + \frac{2\kappa}{\varepsilon}\nabla \times \boldsymbol{\omega} - \frac{\mu + \kappa}{k}\mathbf{v}, \\ &0 = (\alpha + \delta - \varsigma)\nabla <\nabla \cdot \boldsymbol{\omega}> + (\delta + \varsigma)\Delta\boldsymbol{\omega} + 2\kappa\nabla \times \mathbf{v} - 4\kappa\boldsymbol{\omega}, \end{aligned}$$

where $k$ is the permeability and $\varepsilon$ is the porosity of the porous medium. $<\nabla \cdot \boldsymbol{\omega}>$ is the volume-averaged divergence, which can be ignored since the spin field is divergence free for the considered cell geometry. A detailed discussion of the filtration equations for the more general case of media with variable porosity can be found in [35].

Using the aforementioned vector equality the governing equations for the porous region can be presented as ($a < r < b$)

$$\nabla \cdot \mathbf{v}_1 = 0,$$
$$-\left(\frac{\mu}{\varepsilon}+\frac{\kappa}{\varepsilon}\right)\nabla \times \nabla \times \mathbf{v}_1 + \frac{2\kappa}{\varepsilon}\nabla \times \boldsymbol{\omega}_1 - \frac{\mu+\kappa}{k}\mathbf{v}_1 = \nabla P_1, \quad (2)$$
$$-(\delta+\varsigma)\nabla \times \nabla \times \boldsymbol{\omega}_1 + 2\kappa \nabla \times \mathbf{v}_1 - 4\kappa \boldsymbol{\omega}_1 = 0.$$

If the following non-dimensional variables and values are used

$$\tilde{r} = \frac{r}{b}, \quad \ell = \frac{a}{b}, \quad m = \frac{c}{b}, \quad \tilde{\mathbf{v}} = \frac{\mathbf{v}}{U}, \quad \tilde{\boldsymbol{\omega}} = \frac{\omega b}{U}, \quad \tilde{P} = \frac{Pb}{\mu U}, \quad (3)$$

the non-dimensional forms of systems (1) and (2) will be correspondingly

$$\tilde{\nabla} \cdot \tilde{\mathbf{v}}_2 = 0,$$
$$-(\mu+\kappa)\tilde{\nabla} \times \tilde{\nabla} \times \tilde{\mathbf{v}}_2 + 2\kappa \tilde{\nabla} \times \tilde{\boldsymbol{\omega}}_2 = \mu \tilde{\nabla} \tilde{P}_2, \quad (4)$$
$$-\frac{\delta+\varsigma}{b^2}\tilde{\nabla} \times \tilde{\nabla} \times \tilde{\boldsymbol{\omega}}_2 + 2\kappa \tilde{\nabla} \times \tilde{\mathbf{v}}_2 - 4\kappa \tilde{\boldsymbol{\omega}}_2 = 0,$$

and

$$\tilde{\nabla} \cdot \tilde{\mathbf{v}}_1 = 0,$$
$$-\frac{\mu+\kappa}{\varepsilon}\tilde{\nabla} \times \tilde{\nabla} \times \tilde{\mathbf{v}}_1 + \frac{2\kappa}{\varepsilon}\tilde{\nabla} \times \tilde{\boldsymbol{\omega}}_1 - \frac{\mu+\kappa}{k}b^2\tilde{\mathbf{v}}_1 = \mu \tilde{\nabla} \tilde{P}_1, \quad (5)$$
$$-\frac{\delta+\varsigma}{b^2}\tilde{\nabla} \times \tilde{\nabla} \times \tilde{\boldsymbol{\omega}}_1 + 2\kappa \tilde{\nabla} \times \tilde{\mathbf{v}}_1 - 4\kappa \tilde{\boldsymbol{\omega}}_1 = 0.$$

Systems (4) and (5) contain several dimensional constants, which can be combined in three non-dimensional parameters. The dimensions of the viscosities $\mu$ and $\kappa$ are the same, so, the parameter $N^2 = \kappa/(\mu+\kappa)$ introduced in [36] is non-dimensional and called the micropolarity number or the coupling number. Another non-dimensional parameter of the micropolar liquid is the scale factor $L^2 = \dfrac{\delta + \varsigma}{4\mu b^2}$ [36], which represents the relation between the micro scale of the problem $(\delta + \varsigma)/\mu$ and the macro scale $b$. The third parameter $\sigma = b/\sqrt{k}$ characterizes the specifics of the filtration part of the flow. It represents the ratio of the macro scale of the cell $b$ to the micro scale of the porous layer $\sqrt{k}$. With these three non-dimensional parameters introduced, systems (4) and (5) take the following forms (tildes are omitted here and hereafter)

$$\nabla \cdot \mathbf{v}_2 = 0,$$
$$-\frac{1}{N^2}\nabla \times \nabla \times \mathbf{v}_2 + 2\nabla \times \boldsymbol{\omega}_2 = \left(\frac{1}{N^2} - 1\right)\nabla P_2, \qquad (6)$$
$$-L^2 \nabla \times \nabla \times \boldsymbol{\omega}_2 + \frac{1}{2}\frac{N^2}{1-N^2}\nabla \times \mathbf{v}_2 - \frac{N^2}{1-N^2}\boldsymbol{\omega}_2 = 0,$$

and

$$\nabla \cdot \mathbf{v}_1 = 0,$$
$$-\frac{1}{N^2}\nabla \times \nabla \times \mathbf{v}_1 + 2\nabla \times \boldsymbol{\omega}_1 - \frac{\varepsilon \sigma^2}{N^2}\mathbf{v}_1 = \varepsilon\left(\frac{1}{N^2} - 1\right)\nabla P_1, \qquad (7)$$
$$-L^2 \nabla \times \nabla \times \boldsymbol{\omega}_1 + \frac{1}{2}\frac{N^2}{1-N^2}\nabla \times \mathbf{v}_1 - \frac{N^2}{1-N^2}\boldsymbol{\omega}_1 = 0.$$

The symmetry of the flow allows to obtain general solutions of systems (6) and (7) in the form $\mathbf{v}_i(r,\theta) = \{u_i(r)\cos\theta;\ v_i(r)\sin\theta;\ 0\}$, $\boldsymbol{\omega}_i(r,\theta) = \{0;\ 0;\ \omega_i(r)\sin\theta\}$, $P_i(r,\theta) = p_i(r)\cos\theta$, $i = 1, 2$. Both system (6) and system (7) reduce to the scalar differential equations requiring twelve conditions for the correct statement of the boundary value problem (BVP).

The no-slip and no-spin conditions on all solid surfaces were essential in the derivation of the filtration equations (2) in [34]. So, it is necessary to set these conditions on the boundary $r = \ell$:

$$u_1(\ell) = 0, \quad v_1(\ell) = 0, \quad \omega_1(\ell) = 0. \tag{8}$$

Among a variety of conditions at the liquid-porous interface the most natural ones from the mechanical point of view is the continuity of all linear and angular velocity components, i.e.

$$u_1(1-0) = u_2(1+0), \quad v_1(1-0) = v_2(1+0), \quad \omega_1(1-0) = \omega_2(1+0). \tag{9}$$

Besides, the continuity of the stress and couple stress tensor components, normal and tangential to the surface $r = 1$ is adopted in this study. The corresponding components of the stress and couple stress tensors in the micropolar liquid for the chosen coordinate system are

$$t_{rr} = (-p(r) + 2\mu u'(r))\cos\theta,$$

$$t_{r\theta} = \left((\mu+\kappa)v'(r) - (\mu-\kappa)\frac{u(r)+v(r)}{r} - 2\kappa\omega(r)\right)\sin\theta,$$

$$m_{r\varphi} = \left((\delta+\varsigma)\omega'(r) - (\delta-\varsigma)\frac{\omega(r)}{r}\right)\sin\theta.$$

The derivation of the filtration equations for micropolar liquids [34] demonstrated that all viscous terms have the coefficients equal to the viscosities of pure liquid divided by the porosity. So, along with effective viscosity $\mu/\varepsilon$, used in filtration models of Newtonian liquids, also $\kappa/\varepsilon$, $\delta/\varepsilon$ and $\varsigma/\varepsilon$ are to be used instead of $\mu, \kappa, \delta, \varsigma$ in the expressions for the stress and couple stress tensor components in the porous region. If relations (3) are applied, the boundary conditions for stresses and couple stresses will take the following non-dimensional form:

$$-p_1(1-0) + \frac{2}{\varepsilon}u_1'(1-0) = -p_2(1+0) + 2u_2'(1+0), \qquad (10)$$

$$\frac{1}{\varepsilon}v_1'(1-0) - \frac{1-2N^2}{\varepsilon}\frac{u_1(1-0)+v_1(1-0)}{1-0} - 2\frac{N^2}{\varepsilon}\omega_1(1-0) =$$
$$= v_2'(1+0) - (1-2N^2)\frac{u_2(1+0)+v_2(1+0)}{1+0} - 2N^2\omega_2(1+0), \qquad (11)$$

$$\frac{1}{\varepsilon}\omega_1'(1-0) - \frac{\phi}{\varepsilon}\frac{\omega_1(1-0)}{1-0} = \omega_2'(1+0) - \phi\frac{\omega_2(1+0)}{1+0}, \qquad (12)$$

with an additional non-dimensional parameter, $\phi = (\delta - \varsigma)/(\delta + \varsigma)$ introduced in equation (12). Due to the non-symmetry of the couple stress tensor, viscosities $\delta$ and $\varsigma$ entered this condition independently and can not be reduced to the parameters *N* and *L*. It is worth noting that parameter $\phi$ arises in the problem for a flow along the axis of a cylindrical cell [30] and does not appear in the case of flow directed perpendicular to the cell axis [31]. So the non-symmetric properties that the micropolar liquid exhibits at the boundary are substantially predefined by the geometry of the flow. The variation interval of $\phi$ is $[-1; 1]$, the case of $\phi = 0$ ($\delta = \varsigma$) being important, as it implies the absence of an explicit dependence of the solution on $\delta$ and $\varsigma$.

Three more boundary conditions should be set at the surface $r = m$. The first of them is the standard continuity condition for the normal component of the linear velocity:

$$u_2(m) = 1. \qquad (13)$$

As the second condition at the outer boundary of the cell four types of conditions known in classical cell models for Newtonian liquids will be used. They are:

Happel's no-stress condition [15], $\left.t_{r\theta}\right|_{r=m} = 0$:

$$v'_2(m-0) - (1-2N^2)\frac{u_2(m-0) + v_2(m-0)}{m-0} - 2N^2\omega_2(m-0) = 0, \quad (14a)$$

Kuwabara's vorticity-free condition [17], $\left.\text{curl}\, \mathbf{v}_2(r,\theta)\right|_{r=m} = 0$:

$$v'_2(m-0) + \frac{u_2(m-0) + v_2(m-0)}{m-0} = 0, \quad (14b)$$

the symmetry of the velocity profile by Kvashnin [20]

$$v'_2(m-0) = 0, \quad (14c)$$

and the condition of the flow uniformity by Cunningham [19]

$$v_2(m-0) = -1. \quad (14d)$$

The last condition at $r = m$ should deal with microrotation or couple stresses. This could be a Happel-type no-couple stress condition $\left.m_{r\varphi}\right|_{r=m} = 0$:

$$\omega'_2(m-0) - \phi\frac{\omega_2(m-0)}{m-0} = 0, \quad (15a)$$

the no-spin condition, which can be regarded as a Kuwabara-type or a Cunningham-type condition:

$$\omega_2(m-0) = 0, \quad (15b)$$

or a Kvashnin-type symmetry of the spin profile:

$$\omega'_2(m-0) = 0. \quad (15c)$$

Various types of slips, say, $\omega_2(m) - \beta\omega_2'(m) = 0$ or $\omega_2(r)|_{r=m} = n\,\mathrm{curl}\,\mathbf{v}_2(r)|_{r=m}$ with parameters $\beta$ and $n$, can also be taken as the boundary conditions. Some of them are discussed in [25] and references therein. One of the recent works dealing with slip conditions both for the linear velocity and microrotation is [37].

Two BVPs for a flow along and across the axis of the cylindrical cell with Happel's no-stress condition and conditions (15a) or (15b) were considered and compared in [30] and [31]. Conditions (15a) and (15c) coincide for the flow perpendicular to the axis of the cylindrical cell. Both investigations showed that the influence of different boundary conditions (15a-15c) on the solution was rather small. The difference of a few percent was obtained for the hydrodynamic permeability of the membrane calculated using the two aforementioned BVP solutions. Therefore, the present work is focused on the comparison of the BVP solutions with boundary conditions (14a-14d), and condition (15a) is taken as the closing one.

## 2. General solution of the problem

The solution of systems (6) and (7) was obtained totally analytically using the following procedure. The curl operator applied to the momentum equation in system (6) gives

$$\nabla \times \nabla \times \boldsymbol{\omega}_2 = \frac{1}{2N^2} \nabla \times \nabla \times \nabla \times \mathbf{v}_2. \qquad (16)$$

Substituting this relation in the moment of momentum equation of system (6) results in the following expression for $\boldsymbol{\omega}_2$

$$\boldsymbol{\omega}_2 = \frac{1}{2}\nabla \times \mathbf{v}_2 - \frac{L^2}{2N^2}\left(\frac{1}{N^2} - 1\right)\nabla \times \nabla \times \nabla \times \mathbf{v}_2. \qquad (17)$$

Expression (17) substituted into equation (16) reduces it to

$$\nabla \times \nabla \times \mathbf{z}_2 + \frac{N^2}{L^2} \mathbf{z}_2 = \mathbf{0}, \qquad (18)$$

where $\mathbf{z}_2 = \nabla \times \nabla \times \nabla \times \mathbf{v}_2$. Due to the symmetry of the flow, which allowed separation of variables for all the unknown functions, such separation can also be done for $\nabla \times \mathbf{v}_2$ and $\mathbf{z}_2$. For $\nabla \times \mathbf{v}_2$ it looks like $\nabla \times \mathbf{v}_2 = \{0; 0; y_2(r)\sin\theta\}$, where

$$y_2(r) = v_2'(r) + \frac{u_2(r) + v_2(r)}{r}; \qquad (19)$$

for $\mathbf{z}_2$ it reads $\mathbf{z}_2 = \{0; 0; z_2(r)\sin\theta\}$, where

$$z_2(r) = -y_2''(r) - \frac{2y_2'(r)}{r} + \frac{2y_2(r)}{r^2}. \qquad (20)$$

Separation of variables in equation (18) gives the modified Bessel equation $z_2'' + 2\frac{z_2'}{r} - z_2\left(\frac{2}{r^2} + \frac{N^2}{L^2}\right) = 0$. Its solution is $z_2(r) = (\tilde{C}_1 I_{3/2}(Nr/L) + \tilde{C}_2 K_{3/2}(Nr/L))/\sqrt{r}$, where $I_{3/2}(\xi)$, $K_{3/2}(\xi)$ are respectively the modified Bessel and Macdonald functions of the order 3/2, $\tilde{C}_1$, $\tilde{C}_2$ are arbitrary constants. Solving equation (20) with the given function $z_2(r)$ yields $y_2(r)$. The continuity equation combined with equation (19) with the substituted $y_2(r)$ allows us to find both linear velocity components:

$$u_2(r) = \frac{C_1}{r^3} + \frac{C_2}{r} + C_3 + C_4 r^2 + \frac{C_5}{r^{3/2}} I_{3/2}\left(\frac{N}{L}r\right) + \frac{C_6}{r^{3/2}} K_{3/2}\left(\frac{N}{L}r\right),$$

$$v_2(r) = -r u_2'(r)/2 - u_2(r).$$

The angular velocity component is obtained from relation (17) as follows

$$\omega_2(r) = \frac{C_2}{2r^2} - \frac{5}{2}C_4 r - \frac{C_5}{4L^2\sqrt{r}} I_{3/2}\left(\frac{N}{L}r\right) - \frac{C_6}{4L^2\sqrt{r}} K_{3/2}\left(\frac{N}{L}r\right).$$

The presence of an additional member in the momentum equation of system (7) induces modifications in the method of its solution. Applying the same procedure we arrive at the equation for $\nabla \times \mathbf{v}_1$, which cannot be reduced to the equation for $\nabla \times \nabla \times \nabla \times \mathbf{v}_1$ and looks like

$$\nabla \times \nabla \times \nabla \times \nabla \times \nabla \times \mathbf{v}_1 + \left(\frac{N^2}{L^2} + \varepsilon\sigma^2\right)\nabla \times \nabla \times \nabla \times \mathbf{v}_1 + \frac{\varepsilon\sigma^2 N^2}{(1-N^2)L^2}\nabla \times \mathbf{v}_1 = \mathbf{0}.$$

Nevertheless, it allows for separation of variables, which leads to

$$y_1^{(IV)} + \frac{4y_1'''}{r} - \frac{4y_1''}{r^2} - \left(\frac{N^2}{L^2} + \varepsilon\sigma^2\right)\left(y_1'' + \frac{2y_1'}{r} - \frac{2y_1}{r^2}\right) + \frac{\varepsilon\sigma^2 N^2}{(1-N^2)L^2}y_1 = 0, \quad (21)$$

where $y_1(r)$ is defined analogously to $y_2(r)$.

Equation (21) presented as

$$y_1^{(IV)} + \frac{4y_1'''}{r} - \frac{4y_1''}{r^2} - \left(\alpha_1^2 + \alpha_2^2\right)\left(y_1'' + \frac{2y_1'}{r} - \frac{2y_1}{r^2}\right) + \alpha_1^2\alpha_2^2 y_1 = 0, \text{ where constants}$$

$\alpha_1, \alpha_2$ satisfy the system $\begin{cases} \alpha_1^2 + \alpha_2^2 = \dfrac{N^2}{L^2} + \varepsilon\sigma^2, \\ \alpha_1^2\alpha_2^2 = \dfrac{\varepsilon\sigma^2 N^2}{(1-N^2)L^2} \end{cases}$ can be replaced by the Bessel equation $z_1'' + 2\dfrac{z_1'}{r} - z_1\left(\dfrac{2}{r^2} + \alpha_2^2\right) = 0$, where $z_1 = y_1'' + 2\dfrac{y_1'}{r} - y_1\left(\dfrac{2}{r^2} + \alpha_1^2\right)$. Solving these two equations one after another we get $y_1(r) = (\tilde{C}_7 I_{3/2}(\alpha_1 r) + \tilde{C}_8 K_{3/2}(\alpha_1 r) + \tilde{C}_9 I_{3/2}(\alpha_2 r) + \tilde{C}_{10} K_{3/2}(\alpha_2 r))/\sqrt{r}$, where $\tilde{C}_7, \tilde{C}_8, \tilde{C}_9, \tilde{C}_{10}$ are arbitrary constants.

The linear and angular velocity components for Region 1 are then found analogously to the corresponding functions in Region 2, namely:

$$u_1(r) = \frac{C_7}{r^3} + C_8 + C_9 \frac{I_{3/2}(\alpha_1 r)}{r^{3/2}} + C_{10}\frac{K_{3/2}(\alpha_1 r)}{r^{3/2}} + C_{11}\frac{I_{3/2}(\alpha_2 r)}{r^{3/2}} + C_{12}\frac{K_{3/2}(\alpha_2 r)}{r^{3/2}},$$

$$v_1(r) = -ru_1'(r)/2 - u_1(r),$$

$$\omega_1(r) = -\frac{\alpha_1^2}{4N^2}\left(1+\alpha_2^2 L^2\left(1-\frac{1}{N^2}\right)\right)\left(C_9 \frac{I_{3/2}(\alpha_1 r)}{\sqrt{r}} + C_{10}\frac{K_{3/2}(\alpha_1 r)}{\sqrt{r}}\right) -$$

$$-\frac{\alpha_2^2}{4N^2}\left(1+\alpha_1^2 L^2\left(1-\frac{1}{N^2}\right)\right)\left(C_{11}\frac{I_{3/2}(\alpha_2 r)}{\sqrt{r}} + C_{12}\frac{K_{3/2}(\alpha_2 r)}{\sqrt{r}}\right),$$

$\theta$-projections of the equations of motion in systems (6) and (7) give the expressions for $p_2(r)$ and $p_1(r)$ correspondingly.

### 3. Solution of the boundary value problems

The profiles of all linear and angular velocity components as functions of $r$ for $\theta = 0$ are shown in Fig.2 for the obtained solutions with conditions (8-14) and (15a) imposed.

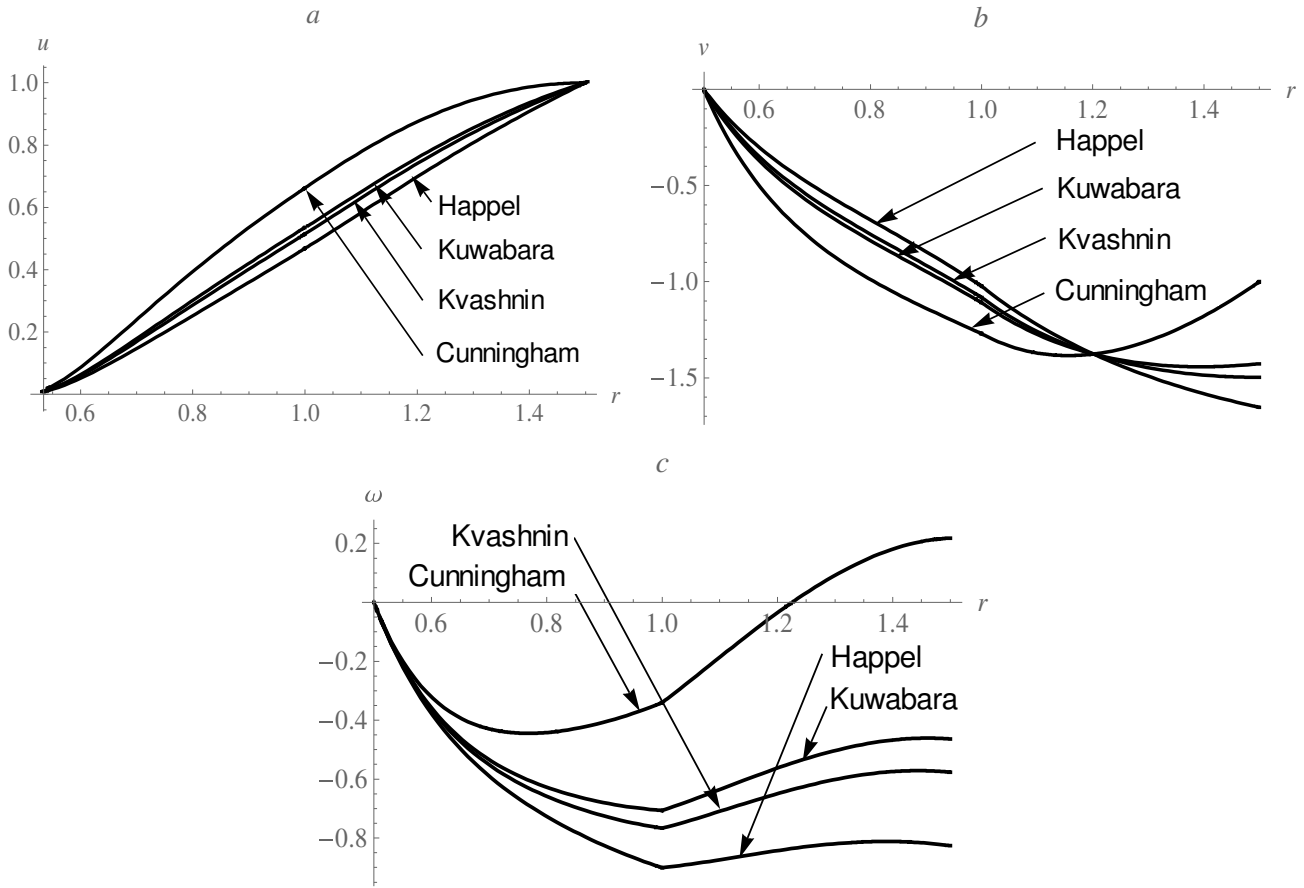

**Fig.2** The radial $u_i(r)$ (a), tangential $v_i(r)$ (b) velocity components and the microrotation $\omega_i(r)$ (c) for BVPs differed by the conditions (14a-14d)

All of the curves in Fig.2 are plotted in non-dimensional units (3) for the following values of parameters: $\ell = 0.5$, $m = 1.5$, $N = 0.5$, $L = 0.2$, $\phi = 0.5$, $\varepsilon = 0.75$ and $\sigma = 3$. These values will also be used in the hydrodynamic permeability investigation. $\ell = 0.5$, $m = 1.5$ correspond to the equal thicknesses of the porous and liquid layers, since the porous-liquid interface is located at $r = 1$. The thickness of the porous layer $1 - \ell$ and the thickness of the solid layer $\ell$ are taken to be equal to each other in order to capture the influence of both of them. The position of the outer cell surface corresponds to the liquid layer of the same thickness. It allows us to obtain more or less pronounced dependence of the solution on the conditions at the outer cell surface. The higher the values of $m$, the further the problem is shifted to the limiting case of a singular sphere in a uniform flow.

The chosen parameters $N = 0.5$, $L = 0.2$, $\phi = 0.5$ of the micropolar liquid represent the well-developed micropolarity properties. As the coupling number $N$ represents a fraction of the rotational viscosity in the sum of the dynamic and rotational viscosity, the range of its possible values is [0;1). So, the chosen value of 0.5 corresponds to one and the same order of magnitude for these coefficients, i.e. to the noticeable influence of the microrotational properties of the medium and the skew symmetric part of the stress tensor. The scale factor $L$ contains the characteristic scale of the problem, $b$ along with the viscosity coefficients. The presence of the macroscopic scale in the solution deprives it of the similarity property, and this fact makes micropolar fluids principally different from the Newtonian ones. Despite the fact that the range of possible values for $L$ is $[0; +\infty)$, very high values of this parameter may have no physical sense. For example, $L > 1$ may be treated as a liquid with structural elements of the order of the whole flow region. On the other hand, one cannot interpret parameter $L$ literally as the ratio of the micro and macro scales of the problem, as no physical particles are actually present in the micropolar liquid. Fig.2 was plotted with $L = 0.2$, which is one of the most reasonable average values for this parameter. The chosen value of $\phi = 0.5$ is far from its possible limiting values, so it allows taking into account the non-symmetry of the couple stress tensor and clarifying its influence on the solution.

Medium values of $\varepsilon = 0.75$ and $\sigma = 3$ are characteristic for a porous medium with neutral properties. It is worth mentioning that the Brinkman equation has been derived for the case of highly porous media ($\varepsilon > 0.6$), so low porosity values are not allowed for its application. Parameter $\sigma$ relates the size of the cell core, *b* and the so-called Brinkman radius $\sqrt{k}$. The latter may be interpreted as the characteristic scale of the filtration flow. So, low values of $\sigma$ correspond to a porous layer almost transparent for the flow. High values of $\sigma$ are attributed to an impermeable porous layer. The chosen value of this parameter is far from both these limiting cases and is associated with a well-developed filtration flow in the porous layer of the cell.

As one can see, the differences in the boundary conditions (14a-14d) relatively weakly affect the radial velocity component and substantially influence the variation of the microrotation velocity (Fig.2c); for the Cunningham condition, even the microrotation direction changes. Anyway, this effect will not arise for sufficiently high values of *m*, i.e. for thick liquid layers.

The most interesting feature of the velocity profiles concerns the behavior of the tangential component (Fig.2b): all the curves have one mutual intersection point. This fact implies the existence of some spherical surface approximately in the middle of the cell liquid layer, where the tangential velocity component of the flow has one and the same value for all conditions at the outer boundary. Moreover, this property holds for any values of all the parameters, both geometrical and mechanical. It is also true for Newtonian liquids and simple solid-liquid cells without a porous layer. The position of the intersection point depends only on the geometrical characteristics of the cell.

Another peculiarity of the presented velocity profiles is the corner points in Fig.2c. These jumps of the microrotation velocity derivatives follow from the difference between the viscosity coefficients for the free micropolar liquid and the corresponding effective viscosities in the porous region. They differ by a factor of $\varepsilon$. Another parameter responsible for this effect is the non-symmetry of the couple stress tensor, expressed by the coefficient $\phi$. In the particular case of $\varepsilon = 1$, $\phi = 0$ all microrotation velocity curves will be smooth; for the micropolar liquid with arbitrary properties this will not be the case.

## 4. Results and discussion

The proposed model and the obtained solution allow for investigation of the membrane hydrodynamic permeability $L_{11}$ which is important for applications. This coefficient is defined as $L_{11} = \dfrac{U}{F/V}$, where the denominator represents the cell pressure gradient, $F$ is the force that the flow exerts on the particle, $V = \dfrac{4}{3}\pi c^3$ is the volume of the cell. The integration of the stresses over the outer surface of the porous layer gives the force $F$

$$F = \oiint_S (t_{rr} \cos\theta - t_{r\theta} \sin\theta) ds.$$

The hydrodynamic permeability takes the following non-dimensional form:

$$L_{11} = \dfrac{4\pi m^3 / 3}{2\pi \int_0^\pi \left( [-p_2 + 2u'_2]\cos^2\theta - \dfrac{1}{1-N^2}\left[ v'_2 - (1-2N^2)\dfrac{u_2+v_2}{r} - 2N^2\omega_2 \right]\sin^2\theta \right)\sin\theta d\theta} \stackrel{T}{=} -\dfrac{m^3}{3C_2}.$$

he results of the parametric study for $L_{11}$ are presented below in the graphical form. The hydrodynamic permeability depends on six parameters: three characteristics of the micropolar liquid, two characteristics of the porous layer and one characteristics of the cell geometry, called active porosity, i.e. the porosity of the membrane as a whole. Each plot shows the dependence of $L_{11}$ on one of the listed parameters for all of the four considered BVPs.

Fig.3 demonstrates the dependence of $L_{11}$ on the micropolarity number $N$. One can see a substantial difference between the curves, corresponding to the BVPs with Happel's, Kuwabara's, Kvashnin's and Cunningham's conditions. This result is non-trivial, as the differences between the velocity components shown in Fig.2 are rather small, so one would expect an additional smoothing effect of the boundary conditions on such integral characteristics of the flow as $L_{11}$. Besides, it follows from Fig.3 that the hydrodynamic permeability is rather sensitive to the variation of N. Maximal values of $L_{11}$ are reached at $N \to 0$ and correspond to the non-polar limit, investigated in [11]. The increase of N implies the intensification of the microrotational effects and the growth of the microrotational viscosity. It leads, in turn, to higher values of the drag force and, consequently, to the diminishing of $L_{11}$. Nevertheless, the limiting case of $N = 1$ can not physically be reached.

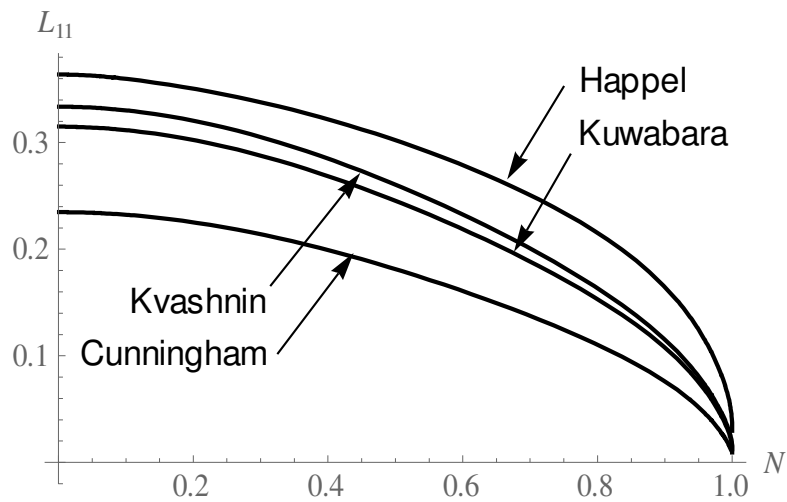

**Fig.3** Variation of hydrodynamic permeability with coupling parameter N for different BVPs

Fig.4 depicts hydrodynamic permeability versus the scale parameter L. A clear difference in the values of $L_{11}(L)$ is again observed for the considered BVPs. Meanwhile, the relative variation of the hydrodynamic permeability for each BVP does not exceed 25%. $L_{11}$ reaches its maximum value at $L = 0$, where the effect of the angular viscosities of the liquid vanishes. With the growth of L the influence of the liquid microstructure increases and the value of $L_{11}$ diminishes, the character of this behavior being independent on the considered types of BVPs. The peculiarity of all the plots for $L_{11}(L)$ is the existence of the asymptotes at $L \to \infty$. As seen in Fig.4, the curves reach their asymptotic values so quickly that the magnitudes of the scale factor $L \sim 1 \div 3$ can be treated as infinite. The asymptotical values of $L_{11}$ are defined by the magnitude of parameter $N$.

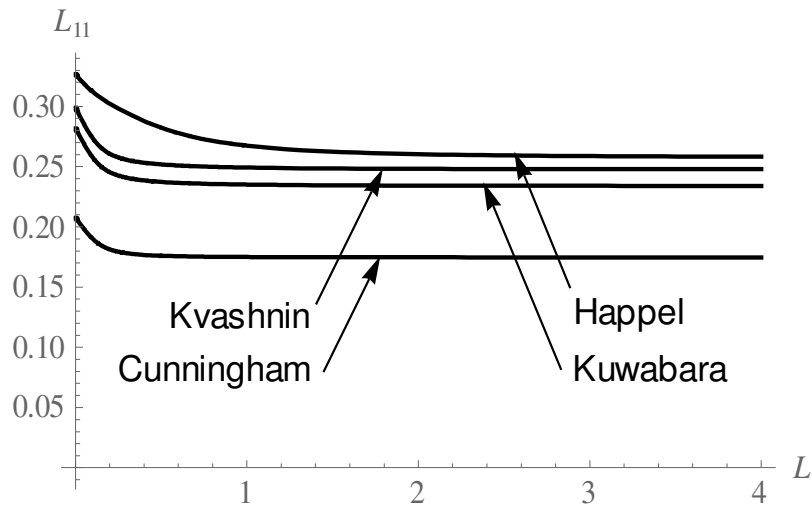

**Fig.4** Variation of hydrodynamic permeability with scale parameter $L$ for different BVPs

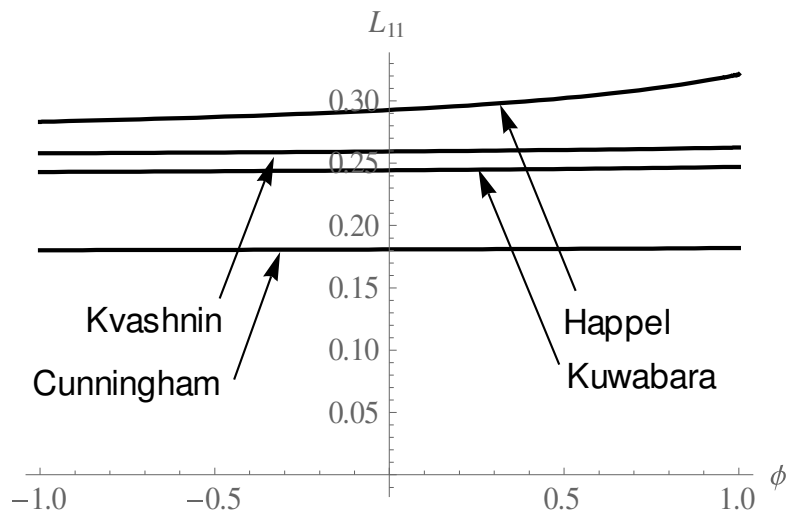

**Fig.5** Variation of hydrodynamic permeability with parameter $\phi$ for different BVPs

The influence of the third parameter of the micropolar liquid, $\phi$ on the hydrodynamic permeability is shown in Fig.5. This parameter enters respectively two boundary conditions for Happel's BVP and one boundary condition for the rest BVPs. As a consequence, it disturbs the behavior of $L_{11}$ for Happel's statement of the problem most of all. The deviation of $L_{11}(\phi)$ from constant for three other BVPs is hardly noticeable as it is seen in Fig.5. And the total variation of $L_{11}(\phi)$ values is about 10% of its average value for Happel's BVP. So in comparison with the influence of parameters $N$ and $L$ on the hydrodynamic permeability, the effect of $\phi$ is almost negligible. It means that the asymmetric properties of the micropolar liquid is not crucial for the evaluation of such integral characteristics of the flow through the membrane, as $L_{11}$.

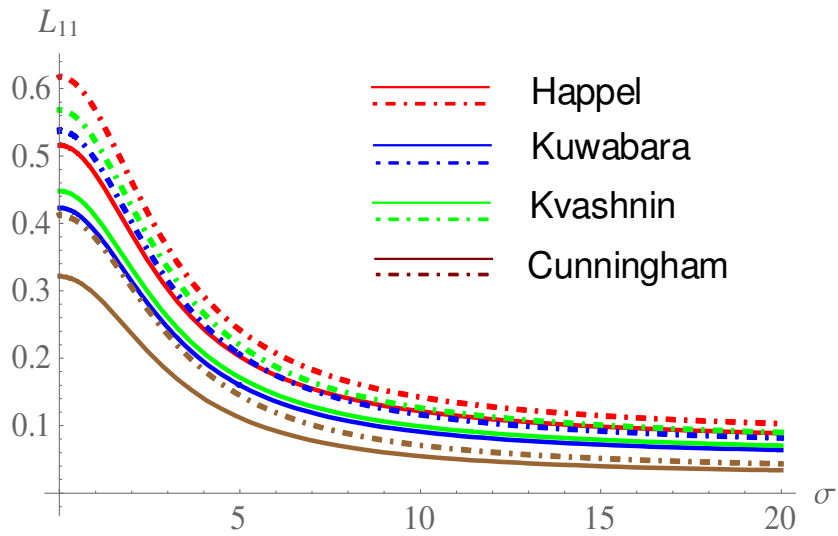

**Fig.6** Variation of hydrodynamic permeability with parameter σ for different BVPs and liquids: micropolar - solid lines, Newtonian - dot-dashed lines

By definition, the permeability parameter σ is inversely proportional to the permeability coefficient of the porous medium of the cell. So, the higher the values of σ the lower the hydrodynamic permeability of the membrane, as it is confirmed by Fig. 6. The values of σ less than unity correspond to so highly-permeable porous layer that its presence can be neglected, and the flow can be considered as totally liquid in the domain of $\ell < r < m$. As a consequence, the highest values of $L_{11}$ are observed for all of the considered BVP statements in this case. For higher values of parameter σ its inversed magnitude can be treated as a fraction of porous layer, where the major part of the filtration flow takes place. The thinner this layer, the less will be the hydrodynamic permeability, in limiting case the flow region will be reduced to the liquid layer $1 < r < m$. Thus, the most complicated case of fully developed filtration in the porous layer corresponds to the values of $\sigma \sim 1 \div 10$. The difference in the hydrodynamic permeability corresponding to the micropolar and classical liquid models is seen in this interval. The curves plotted for the same four types of BVP in the framework of Newtonian liquid model are shown in Fig.6 with the dot-dashed lines. They demonstrate the dependence of $L_{11}(\sigma)$ analogous to that of the micropolar model and confirm the results obtained in [11]. For each of the considered BVPs the curve for Newtonian liquid is located higher than the corresponding curve for the micropolar liquid. This is a natural consequence of the additional degrees of freedom and viscosities presence in the micropolar liquid which leads to lower values of the flow velocity and hydrodynamic permeability. Nevertheless, it is worth paying attention to the position of the plot for Cunningham's model for the Newtonian liquid: apart from the least hydrodynamic permeability among the considered types of BVPs it gives lower values even in comparison with the other BVPs for the micropolar liquid. This circumstance somewhat separates Cunningham's condition from the others.

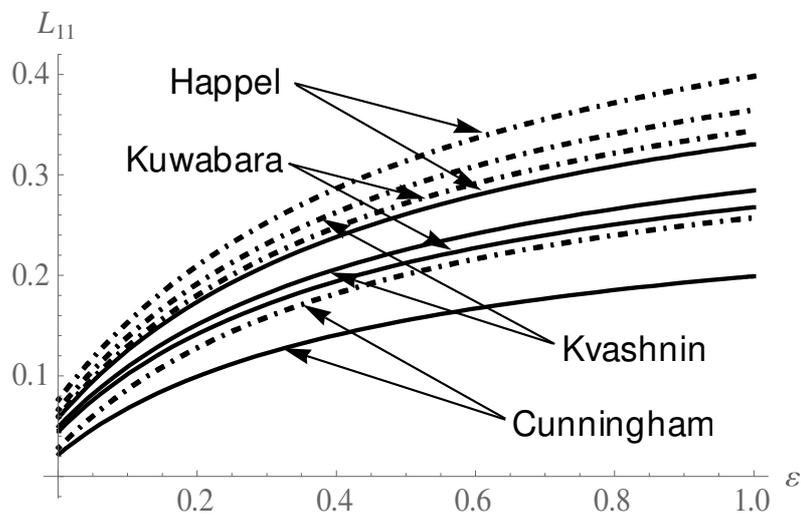

**Fig.7** Variation of hydrodynamic permeability with porosity $\varepsilon$ for different BVPs and liquids: micropolar - solid lines, Newtonian - dot-dashed lines

The porosity $\varepsilon$ of the layer $\ell < r < 1$ is an intrinsic property of the cell, which is weakly connected with the apparent porosity of the membrane as a whole and can hardly be measured in an experiment. The model of the complex porous membrane, developed in the present study is aimed at the simulation of a partially degraded membrane. The solid matrix of such membranes is covered with a partially permeable gel layer which can be considered as the porous Brinkman-type medium [38] with the porosity ε. Anyway, this parameter significantly influences the hydrodynamic permeability of a membrane, as it follows from Fig.7. The whole range of parameter ε variation corresponds to an approximately three-fold growth of $L_{11}$ for both liquid models and all types of BVPs. Nevertheless, this conclusion should be treated cautiously, as the Brinkman-type model was originally designed only for the highly porous media. Quite natural growth of the hydrodynamic permeability with the increase of the porosity ε is observed. Similarly to the dependence of $L_{11}(\sigma)$ the curve $L_{11}(\varepsilon)$ for Cunningham's model in Newtonian liquid demonstrates the values less than for some BVPs for the micropolar liquid.

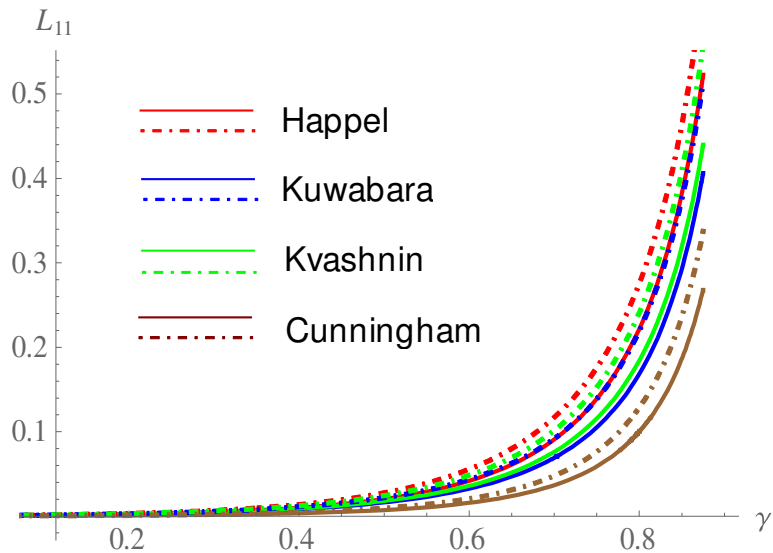

**Fig.8** Variation of hydrodynamic permeability with porosity γ for different BVPs and liquids: micropolar - solid lines, Newtonian - dot-dashed lines

The active porosity of the membrane as a whole, γ is usually measured in experiments. It is defined as fraction of voids filled with the liquid. In the original Happel and Brenner cell models which included only the solid core of radius $b$ and the liquid shell of radius $c$, the porosity γ was calculated as the relative liquid volume of the cell $1-\dfrac{b^3}{c^3}$ and the volume of the inter-cell space was neglected. For the cell construction considered in the present paper, the volume of pores in the layer $\ell < r < 1$ should be added to this magnitude, so it takes the form $\gamma = \dfrac{c^3 - b^3 + \varepsilon(b^3 - a^3)}{c^3} = 1 - \dfrac{1}{m^3} + \varepsilon \dfrac{1 - \ell^3}{m^3}$. For large values of $\ell$ (thin porous layer) the last term can be neglected and the membrane porosity can be calculated simply as $\gamma \approx 1 - 1/m^3$. The dependence of the hydrodynamic permeability on γ defined in this manner is shown in Fig.8 for $\ell = 0.9$. The plot demonstrates the dominating role of the porosity γ in the hydrodynamic permeability determination. The close position of curves to each other and the sharp increase of $L_{11}$ observed for all BVP statements and liquid types imply a secondary role of all these conditions when the dependence of $L_{11}$ on the membrane porosity is studied.

The obtained theoretical dependence of the membrane hydrodynamic permeability on its active porosity was compared with the experimental data of [39]. The experiment considered the flow of a water-ethanol mixture through a nanoporous membrane based on Poly(1-Trimethylsilyl-1-Propyne) for various pressure gradients and mass fraction of the ethanol in the mixture. The membranes were developed using the tape-cast method. In the flow experiments the mass flow rate was measured for each pressure gradient applied to the system. Then the permeability of the membrane was calculated basing on the Darcy law. The dependence of the permeability on the active porosity was obtained for three sets of conditions on the pressure gradient and ethanol concentration. Both these characteristics influence the ability of the membrane to conduct liquid. It is known, that this type of membranes is almost impermeable for a pure water and its permeability increases when the ethanol concentration in the mixture rises. The reason of this effect lies in the hydrophilization of the membrane material by the adsorbed ethanol molecules. This effect was modeled in [39] as the pore opening process and was formalized by the introduction of the effective porosity. The latter was defined as a fraction of pores, opened for the flow. The mass flow rate was evaluated on the basis of the hydrodynamic model that interpreted the filtration as a flow through a set cylindrical pipes of radii subjected to the given distribution law. Then the measured and the theoretical flow rates were equated and the fraction of pores occupied by the liquid was calculated.

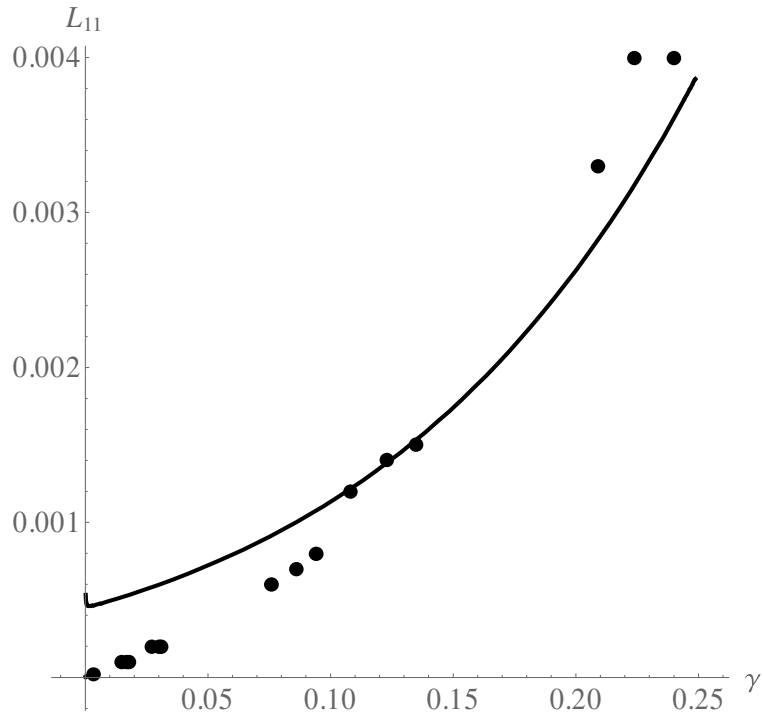

**Fig.9** Comparison of calculated hydrodynamic permeability versus porosity γ with experimental data

In the present study the hydrodynamic permeability $L_{11}$ is non-dimensional, so the experimental data should be driven to the same non-dimensional units. By definition of the hydrodynamic permeability its dimension can be represented as a length squared divided by a viscosity. The characteristic length $b$ used in relations (3) represents the order of the cell size; it was estimated using the simplest orthogonal package of the spherical solid parts of the cells, the voids between them were approximated by spheres. This estimate relates the characteristic scale $b$ to the porosity of the sample and its average pore size which were taken from [39]. In addition, the membrane swelling and its dependence on the mixture composition was taken into account in the calculation of the average pore size as it was offered in [39]. As a result, the value of $b$ turned out to be dependent on the ethanol concentration in the mixture apart from the sample porosity. Each experimental value was divided by the corresponding value of $b^2$ and multiplied by the mixture viscosity. The variation of the latter with the ethanol concentration was also taken into account. The obtained points and the corresponding theoretical curve are shown in Fig.9. The solution for Happel's BVP was taken for plotting although any other curve shown in Fig.8 is applicable as well. The curve in Fig.9 is plotted for the same parameters as in Fig.8 except for $\varepsilon$, which was taken to be equal to 0.6 in order to diminish the value of the hydrodynamic permeability at zero porosity. The model includes the flow through the porous core of the cell, so the calculated velocity and, consequently, the hydrodynamic permeability is non-zero even if the fraction of voids in the membrane is negligibly small ($\gamma \to 0$).

As it is seen from Fig.9, the presented theoretical model gives reasonable agreement with the experiment. The curve represents the character of the dependence, the main trend, the order of magnitude, the convex direction. It overestimates the hydrodynamic permeability for low values of porosity by the reason mentioned before. And it underestimates the value of the hydrodynamic permeability when the porosity increases. The reason of this discrepancy may lie in the fact that the method of the effective porosity evaluation used Newtonian liquid model. It is known that classical model usually gives higher permeability values than the micropolar model.

## 5. Conclusion

The presented study enters the set of works devoted to the development of the cell model technique for filtration flows of micropolar fluids. It includes the most general configuration of the spherical cell, which consists of the solid core, porous shell and liquid envelope. The variation of the thicknesses of these layers allows for the consideration of various limiting cases, namely, the totally solid core ($\ell \to 1$), the totally porous core ($\ell \to 0$), the solid-porous cell ($m \to 1$). Variation of the porous layer parameters allows modeling the membrane permeability in a wide range of values from a negligible one up to the full one, corresponding to a pure matrix without porous coating. The introduced parameters of the permeating liquid represent a wide variety of its properties from classical up to strongly polar ones. The statements of several types of BVPs were given and their analytical solutions were obtained. The expressions for the flow velocity components were presented in the finite form suitable for practical applications.

The hydrodynamic permeability taken as the integral characteristics of the membrane has rather bulk analytical representation, so it was investigated in the computational parametric experiments. They demonstrated the noticeable effect of the considered boundary conditions at the outer surface of the cell on the hydrodynamic permeability. A moderate variation of the hydrodynamic permeability was observed for the whole range of the micropolar parameters of the liquid $N$, $L$, $\phi$ and the characteristics of the porous layer of the cell $\varepsilon$ and $\sigma$. The active porosity of the membrane $\gamma$ gave the strongest effect on the hydrodynamic permeability. The obtained dependence was compared with the experimental data and a reasonable agreement was found.

The interpretation of the experiments remains thus far one of the most difficult challenges in membrane science. The studies of membrane properties are continued even for the commercially available membranes [40]. The presented model is aimed at the improvement and development of the membrane processes understanding. The flexibility of the model allows its application for a wide variety of materials and liquids. The obtained analytical results are suitable for engineering researches. Despite promising advantages, the theory of the micropolar liquids is extremely weakly used in the experiment interpretations today. This paper, hopefully, may serve as a step toward its wider practical application.

**Conflict of Interest**

The author declares that she has no conflict of interest.